\documentstyle[amssymb,epsf,11pt]{article}

\begin{document}

\title{Space-time fractional diffusion equations and asymptotic behaviors of a coupled continuous time random walk model}

\author{ Long Shi$^{1,2}$, ~ Zuguo Yu$^{1}$\thanks{
  Corresponding author, email: yuzg@hotmail.com }, ~ Zhi Mao$^1$, ~ Aiguo Xiao$^1$ ~ and ~Hailan Huang$^1$ \\
 {\small $^{1}$School of Mathematics and Computational Science,
Xiangtan University, Hunan 411105, China.} \\
{\small $^{2}$Institute of Mathematics and Physics, Central South University of Forest and}\\
{\small Technology, Changsha, Hunan 410004, China.}}
\date{}
\maketitle

\begin{abstract}
In this paper, we consider a type of continuous time random walk
model where the jump length is correlated with the waiting time.
The asymptotic behaviors of the coupled jump probability density
function in the Fourier-Laplace domain are discussed. The
corresponding fractional diffusion equations are derived from the
given asymptotic behaviors. Corresponding to the asymptotic
behaviors of the joint probability density function in the
Fourier-Laplace space, the asymptotic behaviors of the waiting
time probability density and the  conditional probability density
for jump length are also discussed.
\end{abstract}


\section{Introduction}

 The continuous time random walk (CTRW) theory, which was
introduced in the 1960s by Montroll and Weiss to describe a walker
hopping randomly on a periodic lattice with the steps occurring at
random time intervals [1], has been applied successfully in many
fields (e.g. the reviews [2-4] and references therein).

In a continuum one-dimensional space, the CTRW scheme is
characterised by a jump probability density function (PDF)
$\psi(x,t)$, which is the probability density that the walker
makes a jump of length $x$ after some waiting time $t$. Let
$P(x,t)$ be the PDF of finding the walker at a given place $x$ and
at time $t$ with the initial condition $P(x,0)=\delta(x)$. A CTRW
process can be described by the following integral equation [3]:
\begin{equation}
\label{(1)}
P(x,t)=\int_{-\infty}^{+\infty}dx'\int_{0}^{t}\psi(x-x',t-t')P(x',t')dt'+\delta(x)\Phi(t),
\end{equation}
where $\Phi(t)=1-\int_{0}^{t}\varphi(\tau)d\tau$ is the probability of not having made a
jump until time $t$ and $\varphi(t)=\int_{-\infty}^{+\infty}\psi(x,t)dx$ is the waiting time PDF.

Fractional diffusion equations (FDEs) arise quite naturally as the
limiting dynamic equations of the CTRW models with temporal and/or
space memories [5]. The asymptotic relation between the CTRW
models and fractional diffusion processes was studied firstly by
Balakrishnan in 1985, dealing with the anomalous diffusion in one
dimension [6]. Later, many authors discussed the relation between
CTRW and FDEs [3-5,7-19]. However, the usual assumption in most of
these works is that the CTRW is decoupled, which means that the
jump lengths and the waiting times are independent. Recently the
coupled CTRW models have attracted more attention [20-25]. Here we
focus on the coupled CTRW models with the jump length correlated
with the waiting time [25], i.e.
$\psi(x,t)=\varphi(t)\lambda(x|t)$, and derive the corresponding
FDEs from the asymptotic behaviors of the waiting time PDF
$\varphi(t)$ and the jump PDF $\psi(x,t)$ in the Fourier-Laplace
space.

This paper is organized as follows. In section 2, we introduce a
space-time fractional diffusion equation which can be obtained
from the standard diffusion equation by replacing the first-order
time derivative and/or the second-order space derivative by a
Caputo derivative of order $\alpha\in(0,2]$ and/or a Riesz
derivative of order $\beta\in(0,2]$, respectively. In section 3,
the asymptotic behaviors of the jump PDF  $\psi(x,t)$ in the
Fourier-Laplace domain are given and the corresponding FDEs are
derived. In section 4, corresponding to the asymptotic behaviors
of the jump PDF $\psi(x,t)$ in the Fourier-Laplace domain, the
asymptotic behaviors of the waiting time PDF $\varphi(t)$ and the
conditional PDF of jump length $\lambda(x|t)$ are discussed. In
section 5, some conclusions are presented.

\section{The space-time fractional diffusion equation}

 We consider a space-time FDE [10]
\begin{equation}
\label{(2)}
_0^CD_t^\alpha u(x,t)=K\frac{\partial^{\beta}u(x,t)}{\partial |x|^{\beta}},\hspace{0.5cm} x\in R, t>0,
\end{equation}
where $u(x,t)$ is the field variable, $K$ is the generalized
diffusion constant and the real paraments $\alpha,\beta$ are
restricted to the range $0<\alpha\leq 2,\ 0<\beta\leq 2$.

In Eq. (2), the time derivative is the Caputo fractional
derivative of order $\alpha$, defined as [26]
\begin{equation}
\label{(3)}
_0^{C}D_t^{\alpha} g(t)
=\left\{\begin{array}{cl}
\frac{1}{\Gamma(n-\alpha)}\int_{0}^{t} \frac{g^{(n)}(\tau)d\tau}{(t-\tau)^{\alpha+1-n}},\hspace{0.5cm}& n-1<\alpha<n,\\
\\
g^{(n)}(t),\hspace{0.5cm}& \alpha=n\in N,
\end{array}\right.
\end{equation}
and the space derivative is the Riesz fractional derivative of order $\beta$, defined as [27]
\begin{equation}
\label{(4)}
\frac{d^{\beta}}{d|x|^{\beta}}f(x)
=\left\{\begin{array}{cl}
\Gamma(1+\beta)\frac{\sin(\beta\pi/2)}{\pi}\int_{0}^{+\infty}\frac{f(x+\xi)-2f(x)+f(x-\xi)}{\xi^{1+\beta}}d\xi,\hspace{0.5cm}& 0<\beta<2,\\
\\
\frac{d^{2}f(x)}{dx^2},\hspace{0.5cm}& \beta=2.
\end{array}\right.
\end{equation}

Let
\begin{equation}
\label{(5)}
\widehat{f}(k)=
{\cal F}\{f(x)\}
=\int_{-\infty}^{+\infty}f(x)e^{ikx}dx
\end{equation}
be the Fourier transform of $f(x)$ and
\begin{equation}
\label{(6)}
\widetilde{g}(s)=
{\cal L}\{g(t)\}
=\int_0^{+\infty}g(t)e^{-st}dt
\end{equation}
be the Laplace transform of $g(t)$.

Now, let us recall the following fundamental formulas about the
Laplace transform of the Caputo fractional derivative of order
$\alpha$ and the Fourier transform of the Riesz fractional
derivative of order $\beta$:
\begin{equation}
\label{(7)}
{\cal L}\{_0^{C}D_t^{\alpha} g(t)\}=s^{\alpha}\widetilde{g}(s)-\sum\limits_{m=0}^{n-1} s^{\alpha-1-m}g^{(m)}(0),\hspace{0.5cm}n-1<\alpha\leq n,
\end{equation}

\begin{equation}
\label{(8)}
 {\cal F}\{\frac{d^{\beta}}{d|x|^{\beta}}f(x)\}=-|k|^{\beta}\widehat{f}(k).
\end{equation}

After applying the formula (7), in the Laplace space, the space-time FDE (2) appears in the form
\begin{equation}
\label{(9)}
s^{\alpha}\widetilde{u}(x,s)-s^{\alpha-1}u(x,0)=K\frac{\partial^{\beta}\widetilde{u}(x,s)}{\partial |x|^\beta}
\end{equation}
for $0<\alpha\leq 1$ and in the form
\begin{equation}
\label{(10)}
s^{\alpha}\widetilde{u}(x,s)-s^{\alpha-1}u(x,0)-s^{\alpha-2}u_{t}(x,0)=K\frac{\partial^{\beta}\widetilde{u}(x,s)}{\partial |x|^\beta}
\end{equation}
for $1<\alpha\leq 2$.

Taking the Fourier transform of Eq.(9) with initial condition
$u(x,0)=\delta(x)$, or of Eq.(10) with initial conditions
$u(x,0)=\delta(x),\ u_t(x,0)=0$, we get
\begin{equation}
\label{(11)}
s^{\alpha}\widehat{\widetilde{u}}(k,s)-s^{\alpha-1}=-K|k|^{\beta}\widehat{\widetilde{u}}(k,s),
\end{equation}
and obtain immediately
\begin{equation}
\label{(12)}
\widehat{\widetilde{u}}(k,s)=\frac{s^{\alpha-1}}{s^\alpha+K|k|^\beta},\hspace{0.5cm}0<\alpha\leq 2, 0<\beta\leq 2.
\end{equation}

{\bf Remark 1}: The fundamental solutions and the asymptotic
solutions of Eq.(2) (containing its special cases) have been
considered in many previous works [10,16,28-34]. In the following
section we will consider two types of asymptotic behaviors of the
jump PDF $\psi(x,t)$ in the Fourier-Laplace space and derive the
corresponding space-time FDEs.

\section{From the coupled CTRW models to FDEs}

 After taking the Laplace transform in the variable $t$ and
the Fourier transform in the variable $x$ of Eq.(1), we get the
following well-known relation [3]
\begin{equation}
\label{(13)}
\widehat{\widetilde{P}}(k,s)=\frac{1-\widetilde{\varphi}(s)}{s}\cdot\frac{1}{1-\widehat{\widetilde{\psi}}(k,s)},
\end{equation}
which is called the Montroll-Weiss equation.

Different types of CTRW processes can be categorised by the existence or non-existence of the characteristic waiting time [3]
\begin{equation}
\label{(14)}
T=\int_{0}^{+\infty}dt \int_{-\infty}^{+\infty}t\psi(x,t)dx,
\end{equation}
and the second moment of the jump length
\begin{equation}
\label{(15)}
\sigma^{2}=\int_{-\infty}^{+\infty}dx \int_{0}^{+\infty} x^{2}\psi(x,t)dt.
\end{equation}

For finite $T$ and $\sigma^2$, the Laplace transform of the waiting time PDF $\varphi(t)$ and the Fourier transform of the jump length PDF $\lambda(x)$ are of the forms
\begin{equation}
\label{(16)}
\widetilde{\varphi}(s)=1-sT+o(s),\hspace{0.5cm} s\rightarrow 0,
\end{equation}
\begin{equation}
\label{(17)}
\widehat{\lambda}(k)=1-\sigma^{2}k^{2}+o(k^2),\hspace{0.5cm} k\rightarrow 0.
\end{equation}

In many applications, one needs to consider long waiting time
and/or long jump length, meaning that the characteristic waiting
time and/or the second moment of the jump length are infinite. It
is natural to generalize Eq. (16) and Eq. (17) to the following
forms [3]:
\begin{equation}
\label{(18 )}
\widetilde{\varphi}(s)=1-A_{\alpha}s^{\alpha}+o(s^{\alpha}),\hspace{0.5cm} s\rightarrow 0,0<\alpha\leq 1,
\end{equation}
and/or
\begin{equation}
\label{( 19)}
\widehat{\lambda}(k)=1-A_{\beta}|k|^{\beta}+o(|k|^{\beta}), \hspace{0.5cm} k\rightarrow 0, 0<\beta\leq 2,
\end{equation}
where $A_{\alpha}$ and $A_{\beta}$ are two positive normal constants.

Therefore, for the decoupled case, in the limit $(k,s)\rightarrow (0,0)$, one has
\begin{equation}
\label{( 20)}
\begin{array}{lll}
\widehat{\widetilde{\psi}}(k,s)
&=& (1-A_{\alpha}s^{\alpha}+o(s^{\alpha}))(1-A_{\beta}|k|^{\beta}+o(|k|^{\beta}))\\
\\
&=& 1-A_{\alpha}s^{\alpha}-A_{\beta}|k|^{\beta}+O(s^{\alpha}|k|^{\beta}).
\end{array}
\end{equation}

In Eq. (20), the term
$1-A_{\alpha}s^{\alpha}-A_{\beta}|k|^{\beta}$ has main influence
on $\widehat{\widetilde{\psi}}(k,s)$ in the limit
$(k,s)\rightarrow (0,0)$. So we can weaken the independent
condition $\psi(x,t)=\lambda(x)\varphi(t)$ and assume  $\psi(x,t)$
has the following form in the Fourier-Laplace domain:
\begin{equation}
\label{(21 )}
\widehat{\widetilde{\psi}}(k,s)= 1-A_{\alpha}s^{\alpha}-A_{\beta}|k|^{\beta}+o(s^{\alpha},|k|^{\beta}),
\end{equation}
which implies that $\psi(x,t)$ is coupled. If
$o(s^{\alpha},|k|^{\beta})=O(s^{\alpha}|k|^{\beta})$, Eq. (21)
reduces to the decoupled case.

Inserting Eq. (18) and Eq. (21) into Eq. (13), in the limit
$(k,s)\rightarrow (0,0)$, we obtain
\begin{equation}
\label{(22)}
\widehat{\widetilde{P}}(k,s)=\frac{s^{\alpha-1}}{s^{\alpha}+K|k|^{\beta}},\hspace{0.5cm} 0<\alpha\leq 1,0<\beta\leq 2,
\end{equation}
where $K=\frac{A_\beta}{A_\alpha}$.

By comparing Eq. (22) with Eq. (12), with the initial condition
$P(x,0)=\delta(x)$, the following space-time fractional equation
is derived immediately:
\begin{equation}
\label{(23)}
_0^{c}D_t^{\alpha} P(x,t)=K\frac{\partial^{\beta}P(x,t)}{\partial |x|^\beta},\hspace{0.5cm}0<\alpha\leq 1, 0<\beta\leq 2.
\end{equation}

{\bf Remark 2}:  We derived Eq. (23) by using the coupled CTRW
model with the asymptotic relations Eqs. (18) and (21). The same
space-time FDE has been also derived using the decoupled CTRW
models in Refs. [14,18,35], where the distribution of the waiting
times and that of the jump lengths are required to be independent
of each other. In Refs. [18,35], the authors showed how the
integral equation for the  CTRW reduces to the space-time
fractional diffusion equation by a properly scaled passage to the
limit of compressed waiting times and jump lengths. Here we extend
their consideration to the coupled case. In Ref. [14], the authors
noted that the same result can be derived by weakening the
independent hypothesis and replacing it with
$\widehat{\widetilde{\psi}}(k,s)\sim 1-s^{\gamma}-|k|^{\beta}$.
But they did not discuss under what conditions one has the above
limiting behavior for the joint distribution $\psi(x,t)$. In the
following section, we will explore the problem and consider a
specific case.

Next, let us extend the asymptotic relation (21) further and
suppose $\psi(x,t)$ has the following form of in the
Fourier-Laplace space:
\begin{equation}
\label{(24)}
\widehat{\widetilde{\psi}}(k,s) \sim 1-A_{\alpha}s^{\alpha}-A_{\beta}\frac{|k|^{\beta}}{s^{\gamma-\alpha}},
\hspace{0.5cm} 0<\alpha\leq 1, \alpha<\gamma\leq 2,0<\beta\leq 2,
\end{equation}
which implies that $\psi(x,t)$ cannot be decoupled in any event.
When $\beta=2$ the asymptotic behavior of
$\widehat{\widetilde{\psi}}(k,s)$ in (24) has been discussed in
Ref. [36].

Inserting Eq. (18) and Eq. (24) into Eq. (13), in the limit
$(k,s)\rightarrow (0,0)$, we obtain
\begin{equation}
\label{(25)}
\begin{array}{lll}
\widehat{\widetilde{P}}(k,s)
&=&\frac{A_{\alpha}s^{\alpha-1}}{A_{\alpha}s^{\alpha}+A_{\beta}|k|^{\beta}s^{\alpha-\gamma}}\\
\\
&=&\frac{s^{\gamma-1}}{s^{\gamma}+K|k|^{\beta}},\hspace{0.5cm} 0<\alpha\leq 1, \alpha<\gamma\leq 2,
\end{array}
\end{equation}
where $K=\frac{A_\beta}{A_\alpha}$.

By comparing Eq.(25) with Eq.(12), we obtain the following
space-time FDE:
\begin{equation}
\label{(26)} _0^{c}D_t^{\gamma}
P(x,t)=K\frac{\partial^{\beta}P(x,t)}{\partial
|x|^\beta},\hspace{0.5cm}0<\alpha\leq 1, \alpha<\gamma\leq 2,
0<\beta\leq 2,
\end{equation}
with the initial condition $P(x,t=0)=\delta(x)$ for $\alpha<\gamma\leq 1$ or the initial conditions $P(x,t=0)=\delta(x), P_{t}(x,t=0)=0$ for $1<\gamma\leq 2$.

\section{The derivation of the asymptotic behaviors of the jump PDF $\psi(x,t)$}

 In this work, we focus on the coupled CTRW model where the
jump PDF $\psi(x,t)$ has the form
$\psi(x,t)=\varphi(t)\lambda(x|t)$, meaning that the jump length
is correlated with the waiting time [25]. In the following, in the
Fourier-Laplace domain, we derive the asymptotic behaviors of
$\widehat{\widetilde{\psi}}(k,s)$ in the limit
$(k,s)\rightarrow(0,0)$ which are introduced in the previous
section.

For the waiting time PDF $\varphi(t)$, we assume in the Laplace space
\begin{equation}
\label{(27)}
\widetilde{\varphi}(s)\sim 1-A_{\alpha}s^{\alpha},\hspace{0.5cm} s\rightarrow 0, 0<\alpha\leq 1.
\end{equation}

For the conditional PDF $\lambda(x|t)$, we assume
\begin{equation}
\label{(28)}
\lambda(x|t)
=\left\{
\begin{array}{lll}
\frac{1}{\sqrt{4\pi g(t)}} \exp(-\frac{x^2}{4g(t)}),\hspace{0.5cm}& if & \beta=2,\\
\\
\frac{1}{(g(t))^{1/\beta}}L_{\beta}(\frac{x}{(g(t))^{1/\beta}}),\hspace{0.5cm}& if &0<\beta<2,
\end{array}\right.
\end{equation}
where $L_{\beta}(x)$ is two-sided L$\acute{e}$vy stable
probability density, defined in Ref. [37]
\begin{equation}
\label{(29)}
L_{\beta}(x)=\frac{1}{2\pi}\int_{-\infty}^{+\infty}\exp(-|k|^{\beta})e^{-ikx}dk
\end{equation}
and $g(t)>0$ is an auxiliary function.
In the Fourier space, we obtain
\begin{equation}
\label{(30)}
\widehat{\lambda}(k|t)=\exp(-g(t)|k|^{\beta}),\hspace{0.5cm}0<\beta\leq 2.
\end{equation}
Then, in the limit $k\rightarrow 0$, we have the asymptotic relation
\begin{equation}
\label{(31)}
\widehat{\lambda}(k|t) \sim 1-g(t)|k|^{\beta}, \hspace{0.5cm} k\rightarrow 0, 0<\beta\leq 2.
\end{equation}

In the Fourier-Laplace space, in the limit $(k,s)\rightarrow(0,0)$
we have
\begin{equation}
\label{(32)}
\begin{array}{lll}
\widehat{\widetilde{\psi}}(k,s)-\widetilde{\varphi}(s)
&=&\int_{0}^{+\infty}dt \int_{-\infty}^{+\infty}\psi(x,t)\exp(ikx-st)dx - \int_{0}^{+\infty}\varphi(t)\exp(-st)dt \\
\\
&=&\int_0^{+\infty}[\widehat{\lambda}(k|t)-1]\varphi(t)\exp(-st)dt \\
\\
&\sim& -|k|^{\beta}\int_0^{+\infty}g(t)\varphi(t)\exp(-st)dt \\
\\
&=& -|k|^{\beta}{\cal L}\{g(t)\varphi(t)\}.
\end{array}
\end{equation}
So
\begin{equation}
\label{(33)}
\widehat{\widetilde{\psi}}(k,s)\sim 1-A_{\alpha}s^{\alpha}-|k|^{\beta}{\cal L}\{g(t)\varphi(t)\}, \hspace{0.5cm} 0<\alpha\leq 1, 0<\beta\leq 2.
\end{equation}

If
\begin{equation}
\label{(34)}
{\cal L}\{g(t)\varphi(t)\} \sim 1-s^{\mu}, \hspace{0.5cm} \mu>0, s\rightarrow 0,
\end{equation}
we can obtain the asymptotic relation
\begin{equation}
\label{(35)}
\widehat{\widetilde{\psi}}(k,s)\sim 1-A_{\alpha}s^{\alpha}-|k|^{\beta}, \hspace{0.5cm} 0<\alpha\leq 1, 0<\beta\leq 2,
\end{equation}
which is the same as Eq. (21).

If
\begin{equation}
\label{(36)}
{\cal L}\{g(t)\varphi(t)\}=\frac{\Gamma(\gamma-\alpha)}{s^{\gamma-\alpha}}, \hspace{0.5cm} 0<\alpha<\gamma, s\rightarrow 0,
\end{equation}
we have
\begin{equation}
\label{(37)}
\widehat{\widetilde{\psi}}(k,s)\sim 1-A_{\alpha}s^{\alpha}-A_{\beta}\frac{|k|^{\beta}}{s^{\gamma-\alpha}},
\hspace{0.5cm} 0<\alpha\leq 1, \alpha<\gamma, 0<\beta\leq 2.
\end{equation}
which is the same as Eq. (24).

 Now we consider the specific case
\begin{equation}
\label{(38)} \varphi(t)\sim t^{-1-\alpha},\hspace{0.5cm}
0<\alpha<1,
\end{equation}
and
\begin{equation}
\label{(39)} g(t)=t^{\gamma}, \hspace{0.5cm} 0<\gamma\leq 2.
\end{equation}
Then
\begin{equation}
\label{(40)}
g(t)\varphi(t)\sim t^{-1-\alpha+\gamma}, \hspace{0.5cm} 0<\alpha<1, 0<\gamma\leq 2.
\end{equation}

If $\gamma<\alpha$, according to the Tauberian theorem [38],  we have
\begin{equation}
\label{(41)}
{\cal L}\{g(t)\varphi(t)\}\sim 1-s^{\alpha-\gamma}, \hspace{0.5cm} 0<\gamma<\alpha<1.
\end{equation}

After taking $\alpha-\gamma=\mu$, Eq. (41) satisfies the condition
Eq. (34). We then obtain the asymptotic relation Eq. (35), and the
corresponding space-time FDE is
\begin{equation}
\label{(42)}
_0^{c}D_t^{\alpha} P(x,t)=K\frac{\partial^{\beta}P(x,t)}{\partial |x|^\beta},\hspace{0.5cm}0<\gamma<\alpha<1,0<\beta\leq 2,
\end{equation}
which implies that the order of time fractional derivative in Eq.
(42) is determined by the parameter $\alpha$ of the waiting time
PDF $\varphi(t)$.

If $\gamma>\alpha$, then $-1-\alpha+\gamma>-1$. Using the Laplace transform formula of power function,  we have
\begin{equation}
\label{(43)}
{\cal L}\{g(t)\varphi(t)\}=\frac{\Gamma(\gamma-\alpha)}{s^{\gamma-\alpha}}, \hspace{0.5cm} 0<\alpha<1, \alpha<\gamma\leq 2.
\end{equation}

It satisfies the condition Eq. (36). So we obtain the asymptotic
relation (37), and the corresponding space-time FDE is
\begin{equation}
\label{(44)}
_0^{c}D_t^{\gamma} P(x,t)=K\frac{\partial^{\beta}P(x,t)}{\partial |x|^\beta},\hspace{0.5cm}0<\alpha<1, \alpha<\gamma\leq 2,0<\beta\leq 2,
\end{equation}
which implies that the order of time fractional derivative in Eq.
(44) is determined by the parameter $\gamma$ of the auxiliary
function $g(t)$.

According to above discussions, we find that for long waiting
time, i.e. $0<\alpha<1$, there exists a competition between the
waiting time PDF $\varphi(t)$ and the auxiliary function $g(t)$ to
decide the order of the time fractional derivative in the
space-time FDEs.

\section{Conclusions}

In this work, we discuss the asymptotic behaviors of the jump PDF
$\psi(x,t)$ in the Fourier-Laplace space in the coupled CTRW model
with $\psi(x,t)=\varphi(t)\lambda(x|t)$. The corresponding
space-time FDEs are derived from the asymptotic behaviors of the
jump PDF $\psi(x,t)$ in the Fourier-Laplace space and the waiting
time PDF $\varphi(t)$ in the Laplace space. We also discuss the
asymptotic behaviors of the conditional PDF of jump length
$\lambda(x|t)$ and show that there exists a competition between
the waiting time PDF $\varphi(t)$ and an auxiliary function $g(t)$
of the  conditional PDF of jump length $\lambda(x|t)$ to determine
the order of the time derivative in the space-time FDE. We also
conclude that when $\beta=2$, the derived FDE Eq.(42) from the
given coupled CTRW model yields subdiffusion. Moreover, FDE (44)
yields subdiffusion for the case of $0<\alpha<\gamma<1$, normal
diffusion for the case of $0<\alpha<\gamma=1$, superdiffusion for
the case of $0<\alpha<1<\gamma\leq 2$.

\section*{Acknowledgements}

 This project was supported by the Natural Science
Foundation of China (Grant no. 11071282 and 10971175), the Chinese
Program for Changjiang Scholars and Innovative Research Team in
University (PCSIRT) (Grant No. IRT1179), the Research Foundation
of Education Commission of Hunan Province of China (grant no.
11A122), the Lotus Scholars Program of Hunan province of China,
the Aid program for Science and Technology Innovative Research
Team in Higher Educational Institutions of Hunan Province of
China. The authors would like to thank Prof. Vo Anh in Queensland
University of Technology for his useful comments and suggestions
to improve this paper.

\end{document}